\documentclass[aps,twocolumn,prl,preprint,groupedaddress,10pt]{revtex4}

\usepackage{graphicx}

\begin{document}


\title{Higher order intercommutations 
in Cosmic String Collisions}


\author{A. Ach\'{u}carro}
\email[]{Achucar@lorentz.leidenuniv.nl}
\affiliation{Instituut-Lorentz for Theoretical Physics, Leiden, The Netherlands}
\affiliation{Departamento de F{\'\i} sica Te\'orica, Universidad del Pa{\'\i}s Vasco UPV-EHU, Bilbao, Spain}
\author{G.J. Verbiest} \email[]{Verbiest@physics.leidenuniv.nl}
\affiliation{Instituut-Lorentz for Theoretical Physics, Leiden, The Netherlands}


\date{\today}

\begin{abstract}

  We report the first observation of multiple intercommutation (more
  than two successive reconnections) of cosmic strings at ultra-high
  collision speeds, and the formation of ``kink trains'' with up to
  four closely spaced left- or right-moving kinks.  We performed a
  flat space numerical study of abelian Higgs cosmic string
  intercommutation in the type-II regime $\beta > 1$ (where $\beta =
  m^2_{scalar} / m^2_{gauge}$) up to $\beta = 64$, the highest value
  investigated to date.  Our results confirm earlier claims that the
  minimum critical speed for double reconnection goes down with
  increasing $\beta$, from $\sim 0.98 c$ at $\beta = 1$ to $\sim 0.86
  c$ for $\beta = 64$. Furthermore, we observe a qualitative change in
  the process leading to the second intercommutation: if $\beta \geq
  16$ it is mediated by a loop expanding from the collision point
  whereas if $1 < \beta \leq 8 $ the previously reported ``loop'' is just
  an expanding blob of radiation which has no topological features and
  is absorbed by the strings. The multiple reconnections are observed
  in the loop-mediated, deep type-II regime $\beta \geq 16$.  Triple
  reconnections appear to be quite generic for collision parameters on
  the boundary between single and double reconnection. For $\beta =
  16$ we observe quadruple events.  They result in clustering of small
  scale structure in the form of ``kink trains''.
  Our findings suggest that, due to the core interactions, the small
  scale structure and stochastic gravitational wave background of 
  abelian Higgs strings in the strongly type-II regime  
  may be quite different from what would be expected from
  studies of Nambu-Goto strings or of abelian Higgs strings in the
  $\beta \approx 1$ regime.

\end{abstract}


\maketitle


Formation of cosmic strings and superstrings is a generic outcome
of cosmological phase transitions\cite{Kibble} and some
inflationary models, both within Grand Unified
theories\cite{Jeannerot,JeannerotRocherSakellariadou} and from branes
\cite{MajumdarDavis,SarangiTye}. There is no evidence of their
existence yet, but strings can lead to a wealth of astrophysical phenomena.
There is an increasing number of surveys and searches looking for
observable signatures such as CMB anisotropies, gravitational lensing,
gravitational radiation, cosmic rays and gamma ray bursts (see
classic reviews \cite{VilenkinShellard, HindmarshKibble} and more
recent updates \cite{CopelandKibble,
  Sakellariadou:2009ev,Achucarro:2008fn}).

A string network is expected to reach a scaling solution in which
statistical properties such as the distance between strings become a
fixed fraction of the horizon size $t$. The
expansion of the Universe pumps energy into the string network --by
increasing the contribution of long strings-- but this is balanced by
energy losses to radiation. If these are efficient enough, the
contribution from the strings to the energy density of the Universe
remains a small, constant fraction of the dominant form of energy
(matter or radiation) and is potentially observable. Radiation is
emitted by oscillating loops, formed when a string self-intersects and
reconnects, and in bursts. The latter are produced by cusps (sections
of the string which acquire near-luminal speeds) and, to a lesser
extent, by kinks created when strings reconnect.  The reconnection, or
{\it intercommutation} of strings is therefore an essential process
that determines and maintains the long-term scaling behaviour of the
string network.

Cosmic string intercommutation was first investigated numerically in
Shellard's pioneering work on global strings \cite{Shellard} and later
by Matzner\cite{Matzner} for type-II abelian Higgs strings with $\beta
= 2$ (using the type-I/type-II terminology from superconductors, to
indicate $\beta<1$ / $\beta>1$ respectively). They pointed out that in
ultra-relativistic collisions there is a critical center-of-mass
velocity $v_c$ (collision angle dependent) beyond which strings pass
through: a loop forms, expanding rapidly from the collision point,
that catches up with the reconnected strings and produces a second
intercommutation. A more recent study \cite{Putter} focussed on double
reconnection, which proceeds differently in type-I and type-II
strings, with the loop only forming in type-II collisions. The angular
dependence of $v_c$ is calculable from the geometry and speeds of the
colliding strings \cite{Putter}.

In the abelian Higgs model, all evidence to date \cite{Matzner,
  Putter} indicates that Abrikosov-Nielsen-Olesen (ANO) strings with
unit winding always reconnect at least once, {\it even in
  ultra-high-speed collisions}, with the possible exception of
near-parallel collisions at very low speed in the type-I regime, where
the attractive force between the strings can make them stick
together\cite{BettencourtKibble}. Theoretical predictions for junction
formation based on the Nambu--Goto
approximation\cite{CopelandKibbleSteer} give an extremely good fit to
numerical simulations \cite{Salmi}.

The simulations in \cite{Putter} had decreasing resolution with
increasing $\beta$, and only explored two values in the type-II regime
($\beta = 8,32$), but the results suggested that the critical velocity
for double reconnection would go down as a function of $\beta$.
This is interesting because one of the distinguishing features of
cosmic superstrings, as opposed to ANO strings, is their low
intercommutation probability $P \sim 10^{-3} - 10^{-1}$
\cite{Jackson:2004zg}, which leads to different scaling properties, in
particular to denser networks (although the effect may be weaker than
the $\rho t^2 \sim 1/P$ one might
expect\cite{AvgoustidisShellard}). However, if the critical velocity
of strongly type-II ANO strings decreases and becomes
comparable to the average velocity of the network, the {\it effective}
intercommutation probability could be much less than one.

While investigating the critical velocity for large $\beta$ we
discovered higher order intercommutation events, that is, particular
collisions in which the strings exchange ends more than twice. These
we describe now.

\section*{Simulations}\label{Simulations}

The abelian Higgs model -the relativistic Ginzburg-Landau model-
involves a complex scalar field $\phi$ and a U(1) gauge field $A_\mu$
with field strength $F_{\mu\nu} = \partial_\mu A_\nu - \partial_\nu
A_\mu$ ($\mu, \nu = 0,1,2,3$). In natural units $\hbar = c = 1$ it is
described by the lagrangian
\[
\mathcal{L} = (\partial_\mu + i e A_\mu)\phi(\partial^\mu - i e A^\mu)\phi^\dagger - \frac{1}{4} F^{\mu\nu} F_{\mu\nu} - \frac{\lambda}{4}(|\phi|^2 - \eta^2)^2
\]
The ground state has $|\phi| = \eta$ and zero magnetic field. There
are two mass scales: scalar excitations have $m_{scalar} = \sqrt
\lambda \eta$ and gauge field excitations, $m_{gauge}=\sqrt 2 e
\eta$. Classically, the only relevant parameter in the dynamics is
their ratio, $\beta = (m_{scalar}/m_{gauge})^2 = \lambda / 2e^2$,
which also characterizes the internal structure of the ANO
vortices\cite{Abrikosov, Nielsen:1973cs}.  In this paper we are
interested in the $\beta > 1$ regime, analogous to a type-II
superconductor, and in this case the vortices have an inner ``scalar''
core of radius $\sim m_{scalar}^{-1}$ in which the scalar field
departs from its vacuum value and vanishes at the center. This is
surrounded by a larger, ``gauge'' core of radius $\sim m_{gauge}^{-1}$
where the magnetic field is non-zero. As a result, the interaction
between parallel type-II strings is repulsive.

We follow the same numerical strategy as in previous studies
\cite{Matzner,Putter}: we use a lattice discretization and place a
superposition of two oppositely moving ANO strings on a three
dimensional lattice. This configuration is evolved using a leapfrog
algorithm. The initial configuration is determined by
two parameters: the center-of-mass speed $v$ of the strings when they
are far apart and the angle $\alpha$ between them (every collision can
be brought to this form by an appropriate
Lorentz transformation \cite{Shellard}).
We also impose  ``freely moving'' boundary conditions:
after each round the fields inside the box are updated using the
equations of motion, and the fields on the boundaries are calculated
assuming the strings move unperturbed and at
constant speeds at the boundaries.
All  simulations shown were done on a $200 \times 200 \times 400$ grid,
with lattice spacing $a = 0.2$
and time steps $\Delta t = 0.02$, so the Courant condition (here
$\Delta t \leq a/\sqrt{3}$) holds.

Our simulations are optimized for the deep type-II regime.
By solving the two-dimensional, static vortex equations one finds,
that in a static straight cosmic string half of the
potential energy in the scalar core is contained within a radius
$\sqrt 2 m_{scalar}^{-1} f(\beta)$, where $f$ is a slowly varying
function with $f(1)= 1, f(64) = 1.4$. Lorentz-contraction gives an
extra factor $\gamma(v)^{-1}$ in the direction of approach, with
$\gamma(v) = 1 / \sqrt{1 - v^2}$.  This is the smallest length scale
that has to be resolved. We fix its size to match the $\beta = 1$
case (without loss of generality we take $\lambda = 2, \ \eta = 1$).
The scalar core is resolved by at least three lattice points up to a center of mass speed of $v \approx 0.94-0.96$, which is indicated explicitly in figure \ref{vcrit_panels}.  We define the gauge field core radius to be larger by a factor $\sqrt{\beta}$. The initial string separation is fixed to five times the gauge core radius .  For $\beta\leq 64$ the gauge cores still do not overlap in the initial configuration.

\section*{Results\label{Results}}

We simulated the collision of cosmic strings at $\beta = 1, 3.9, 4.0,
4.1, 8, 16, 31, 32, 33, 49$ and $64$ for various speeds $v$ and angles
$\alpha$ to find the critical velocity above which the strings
effectively pass through each other. The results for $\beta = 16$ are
shown in figure \ref{vcrit_panels}, other $\beta$ values and a detailed
discussion can be found in ref.\cite{us}.

Our results confirm a downward trend with $\beta$ of the lowest
critical velocity $v_{c, min}$.
Specifically, we
find 
$(\beta, v_{c, min})$ = $(4, 0.92)$, $(8, 0.92)$, $(16, 0.90)$, $(32, 0.88)$, $(49, 0.88)$ and $(64, 0.86)$.

\begin{figure}[!h]
\begin{minipage}{86mm}
\begin{tabular}{cc}
\includegraphics[height = 25mm,width=40mm]{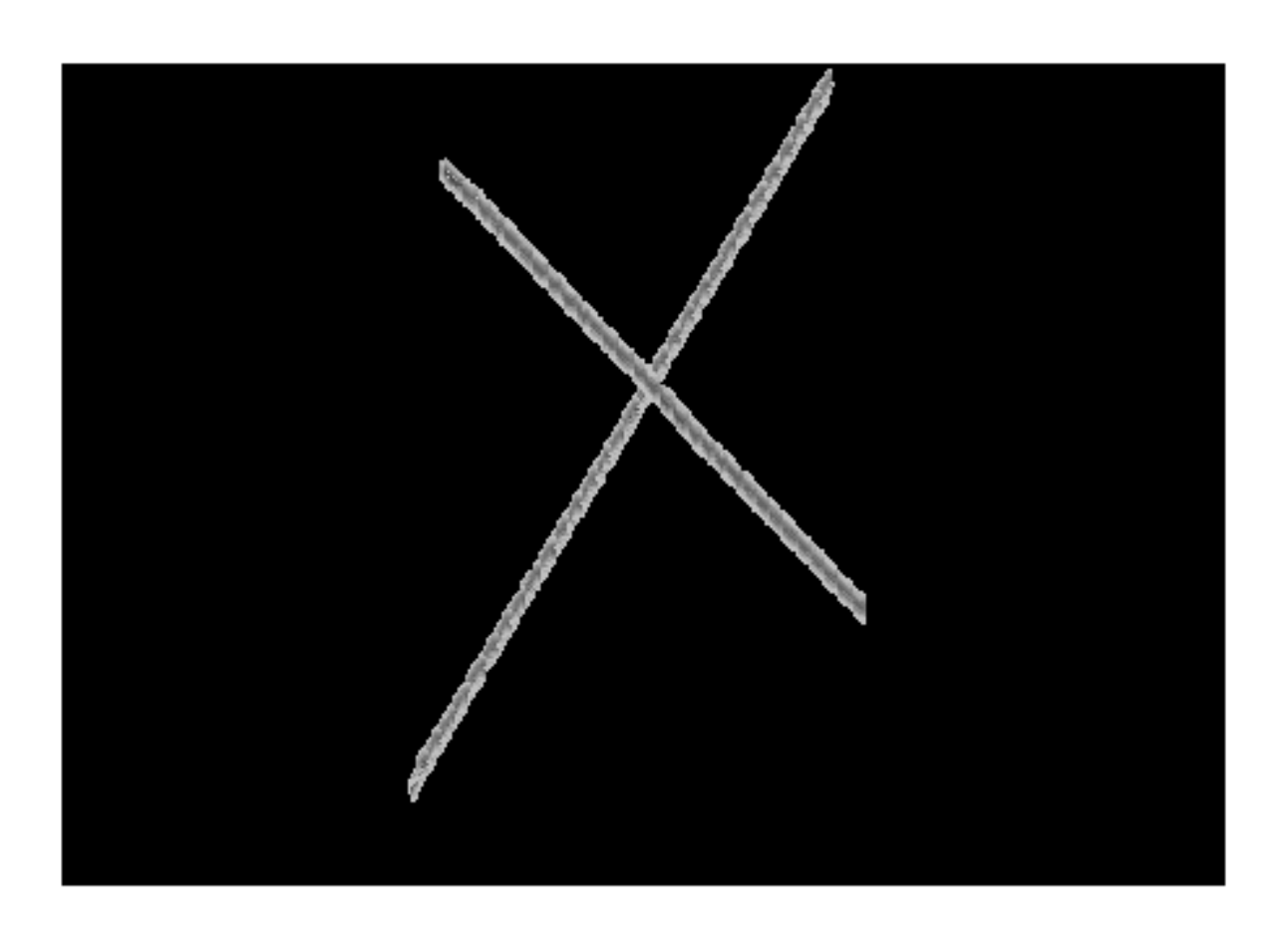} &
\includegraphics[height = 25mm,width=40mm]{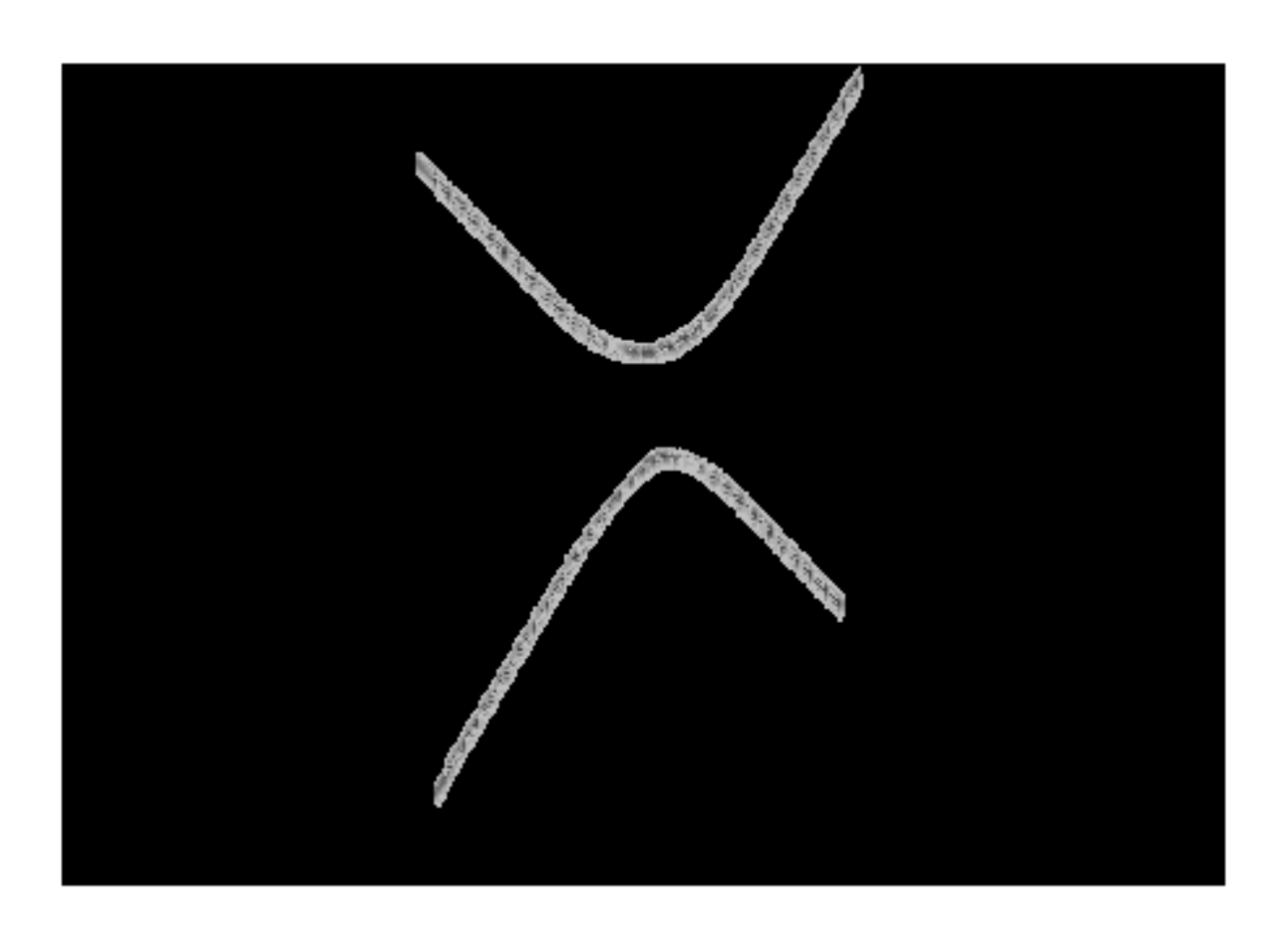} \\
\includegraphics[height = 25mm,width=40mm]{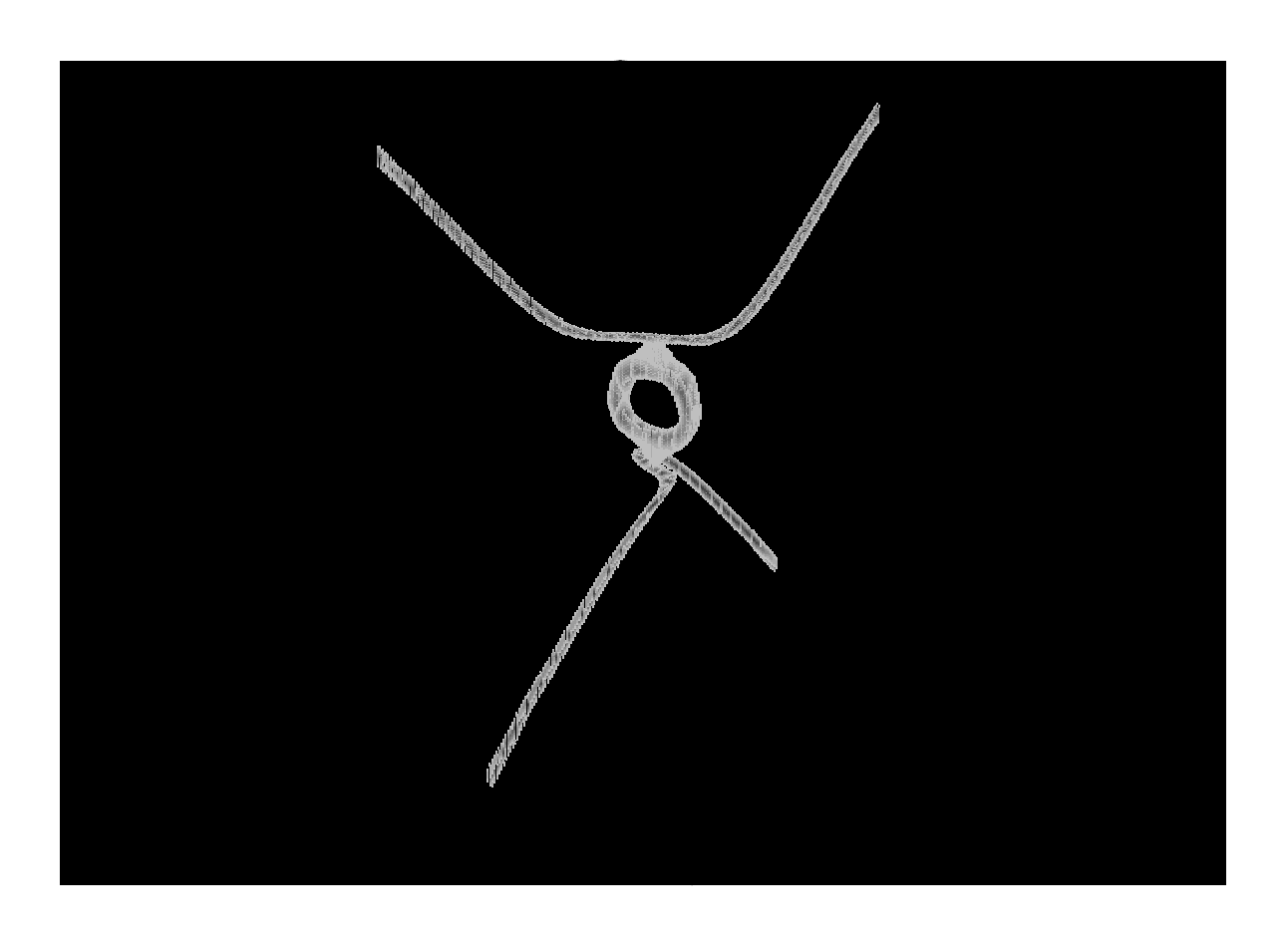} &
\includegraphics[height = 25mm,width=40mm]{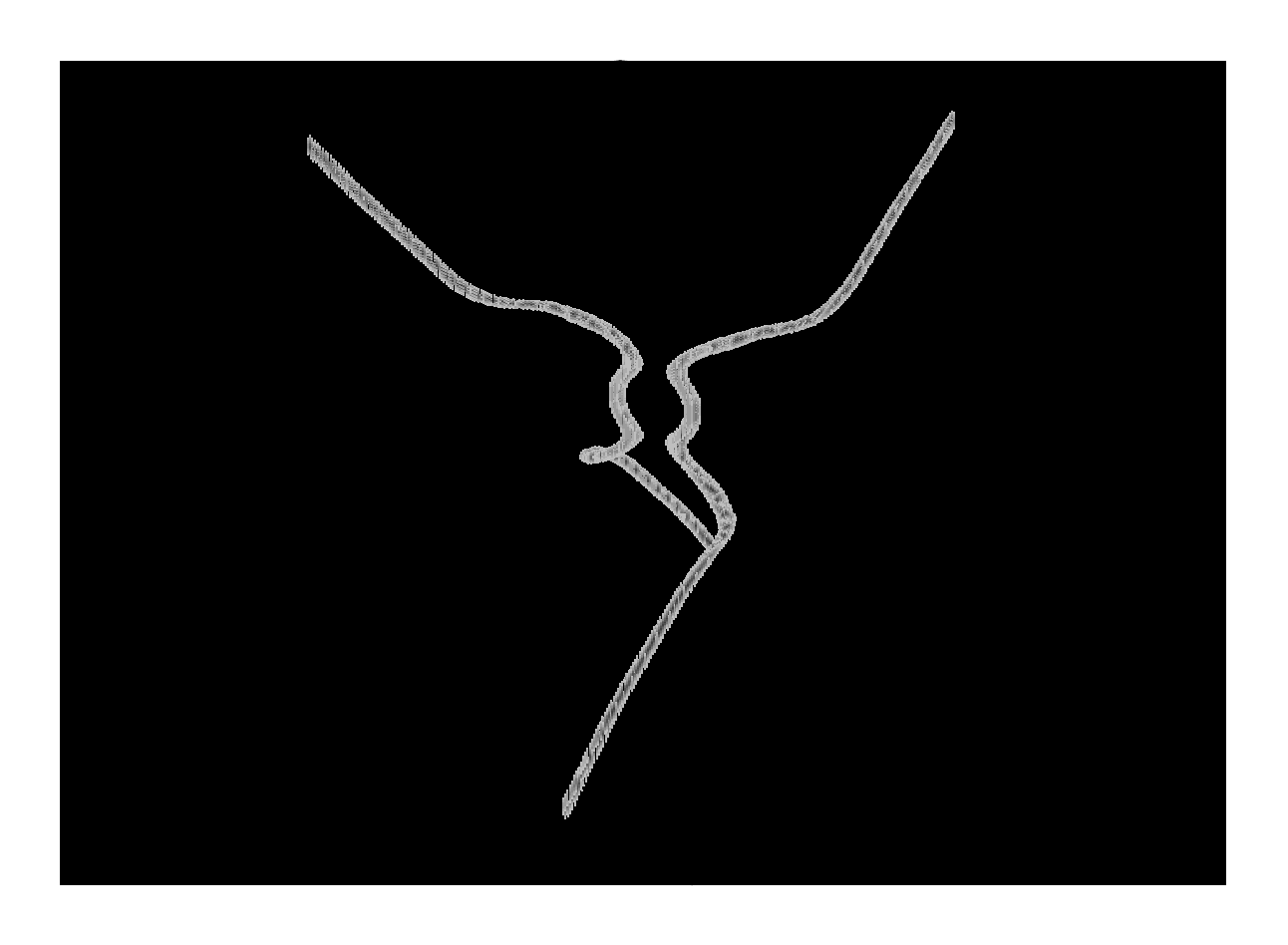} \\
\includegraphics[height = 25mm,width=40mm]{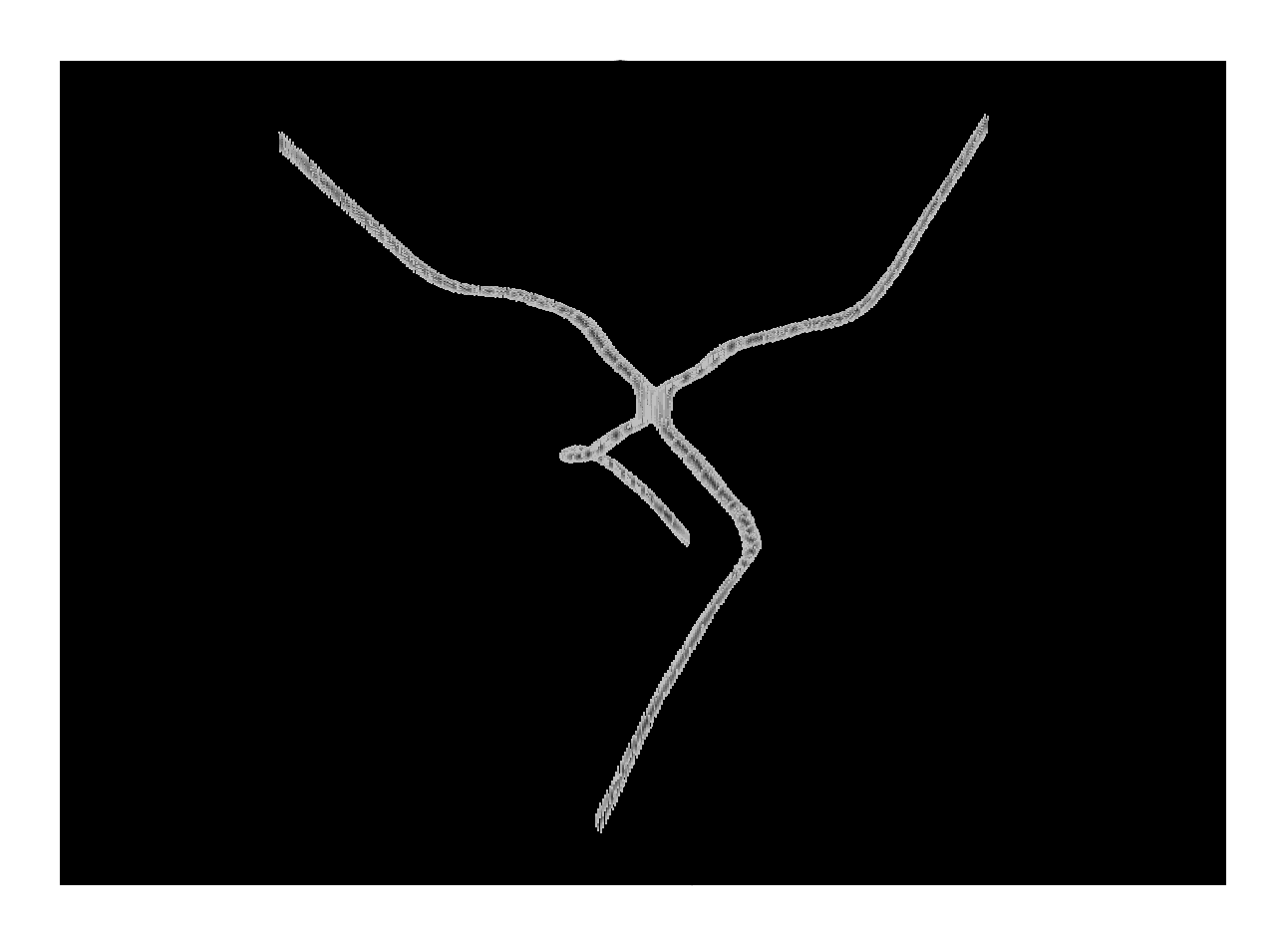} &
\includegraphics[height = 25mm,width=40mm]{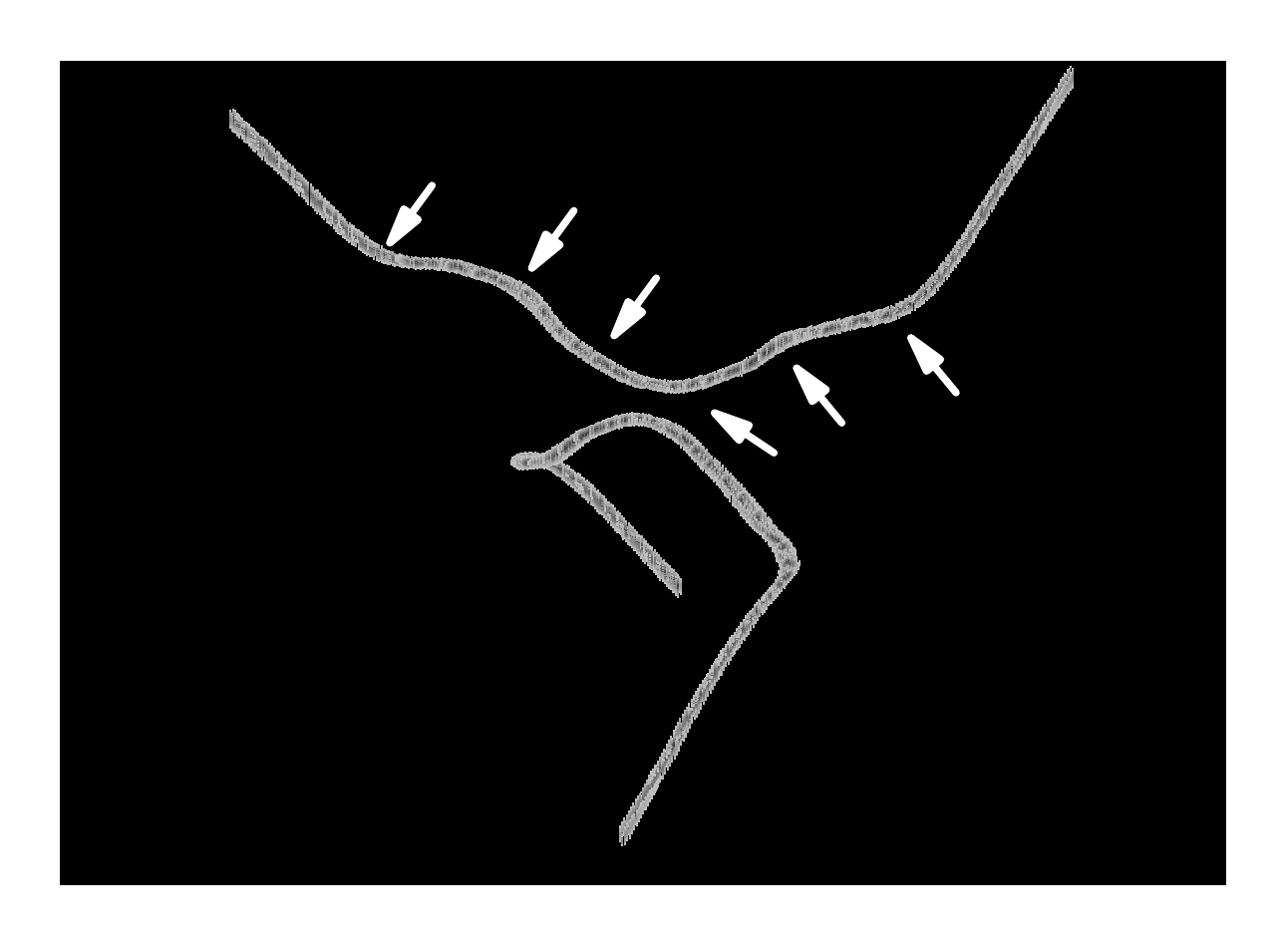} \\
\end{tabular}
  \caption{\label{triple_panels} Scalar field isosurfaces with
    $|\phi|\leq 0.4$. After the first intercommutation a loop emerges
    that catches up with the receding strings (second
    intercommutation), creating two highly curved central
    regions. These move towards each other and produce a third
    intercommutation. Two sets of three closely spaced, left- and
    right-moving kinks are clearly visible on each of the strings
    (indicated by arrows on the upper string). $(\beta = 32, \alpha =
    122.7, v = 0.88)$, $t = 3, 5, 10, 17, 20, 22$}
\end{minipage}
\end{figure}

\begin{figure}[!h]
\begin{minipage}{86mm}
\includegraphics[height = 23mm,width=40mm]{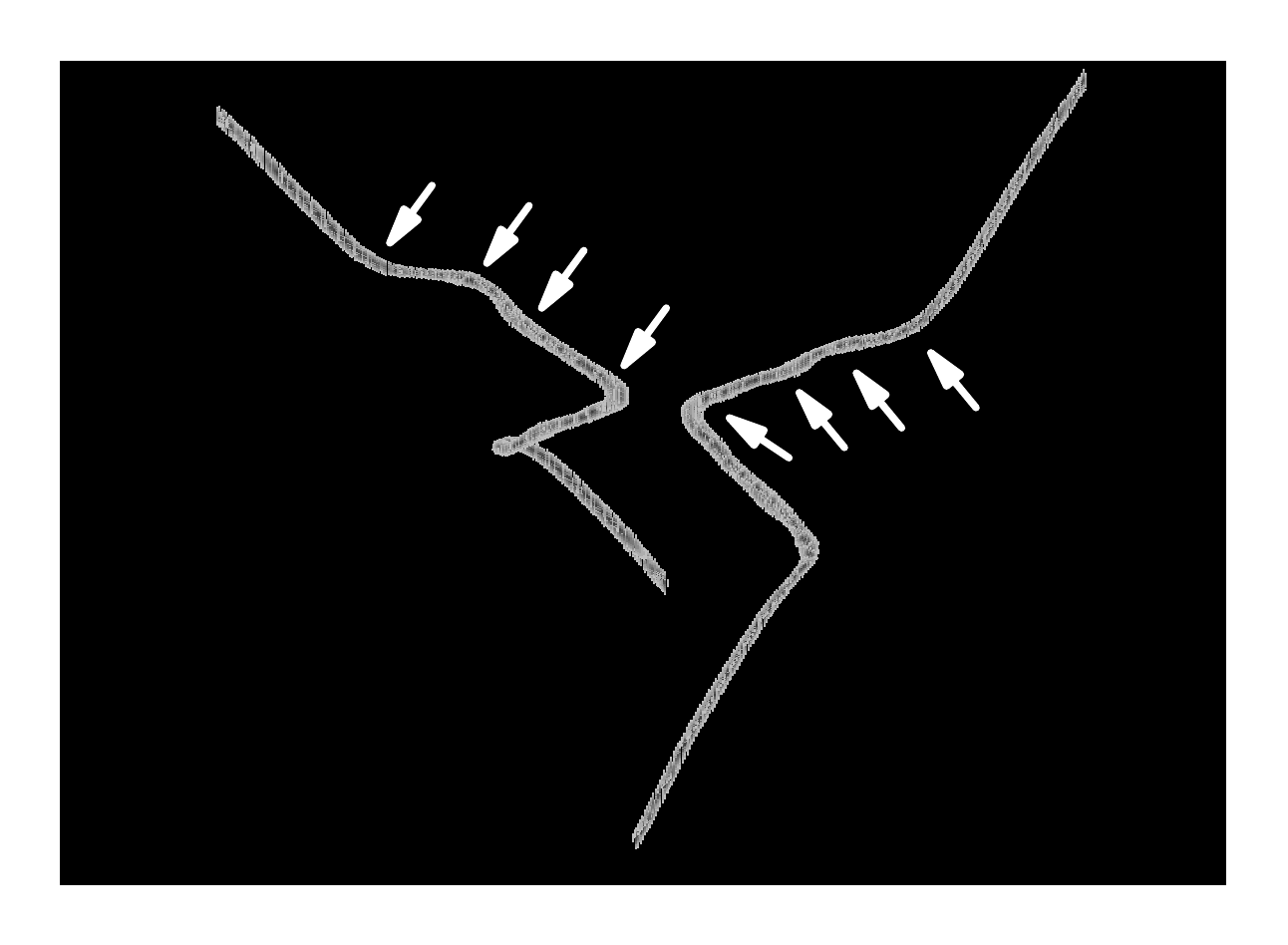}
\caption{
  Typical string configuration after the fourth intercommutation. The arrows show four closely spaced kinks moving up each string; the two corresponding down-moving trains are visible below the collision
point. $(\beta = 16, \alpha = 122.7, v = 0.92)$, same
  isosurfaces, $t = 20$.}
\label{quadruple_final}
\end{minipage}
\end{figure}

A closer look at the interaction process reveals that the strings do
not always intercommute once or twice, as previously observed, but
also three and four times for particular values of the initial speed
$v$ and angle $\alpha$. An odd number of reconnections results in
overall intercommutation of the strings segments, and an even number
in the strings effectively passing through, so we can still speak of a
critical velocity for the strings passing through. However, each
reconnection creates small structure on the strings in the form of a
left- and a right-moving kink. In some cases it is not easy to
distinguish between one and three, or between two and four
reconnections, just by looking at the energy isosurfaces in the
intermediate state. But the resulting kinks are clearly visible in the
final state and can be counted; in case of doubt we use this criterion
(see fig.  \ref{quadruple_final}). Successive reconnections therefore
lead to left- or right-moving ``kink trains'', groups of up to four
closely spaced kinks (one for each intercommutation). The inter-kink
distance within these trains is a few string widths, at formation (see
figs \ref{triple_panels} and \ref{quadruple_final}).

The intercommutation process for type II strings
unfolds as follows. After the collision in which the strings exchange
ends for the first time two things can happen:

\begin{figure}[!h]
\includegraphics[width=86mm]{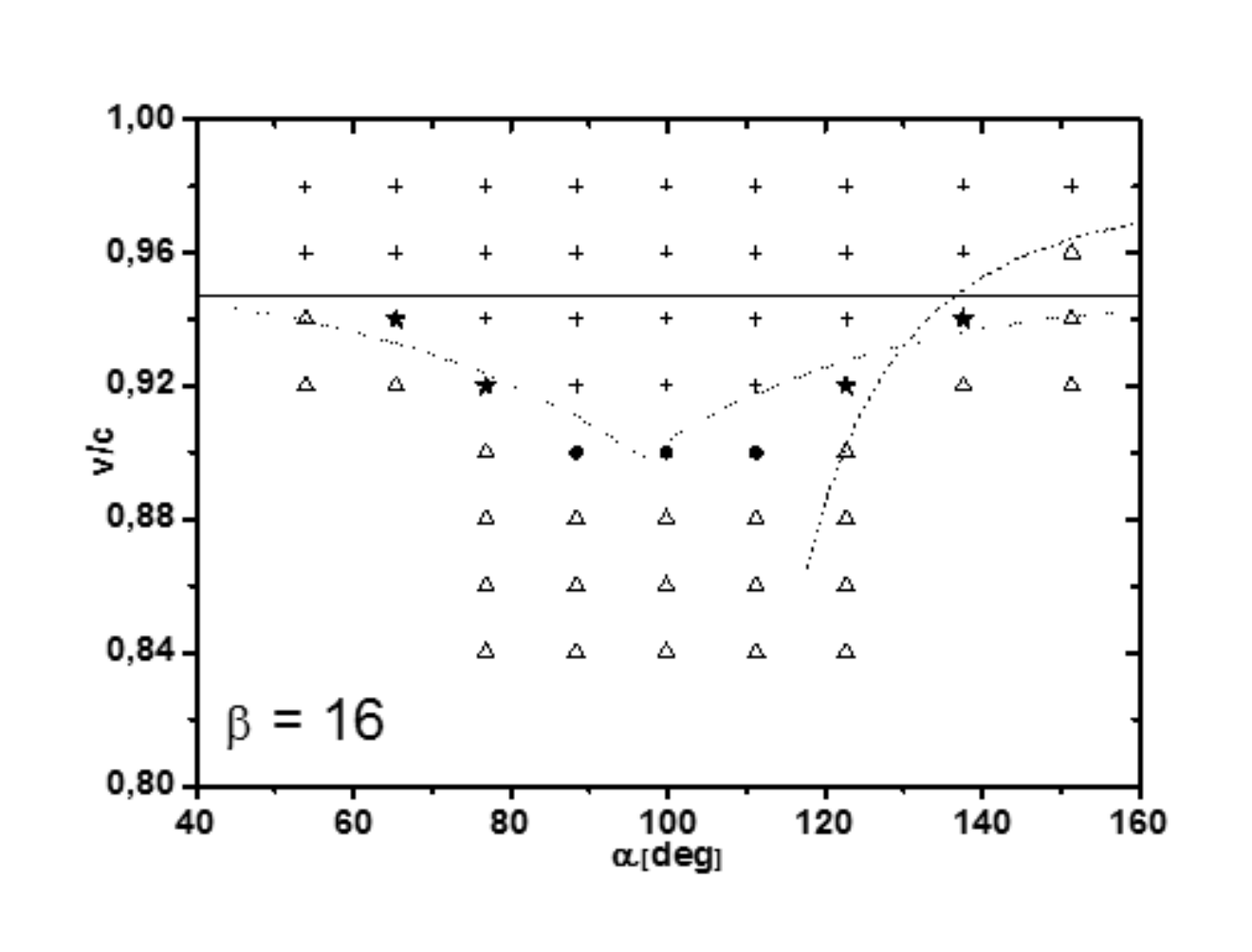}\\
\caption{\label{vcrit_panels} The number of intercommutations for a
  range of collision velocities and angles for $\beta = 16$. The
  symbols $\triangle$, $+$,$\bullet$ and $\star$ stand for 1,2,3 or 4
  intercommutations respectively.  The dotted line is a two-parameter
  fit based on the Nambu--Goto approximation\cite{Putter}. The dashed
  line includes core interaction effects at high collision angle $
  \sim 180^o$ \cite{us}.  Simulations above the horizontal line
  resolve the scalar core size by less than three lattice points and
  are therefore less reliable. Triple and quadruple reconnections
  occur on the boundary betwen single and double reconnections. }
\end{figure}

\begin{figure}[!h]
\begin{minipage}{86mm}
\begin{tabular}{cc}
\includegraphics[height = 23mm,width=40mm]{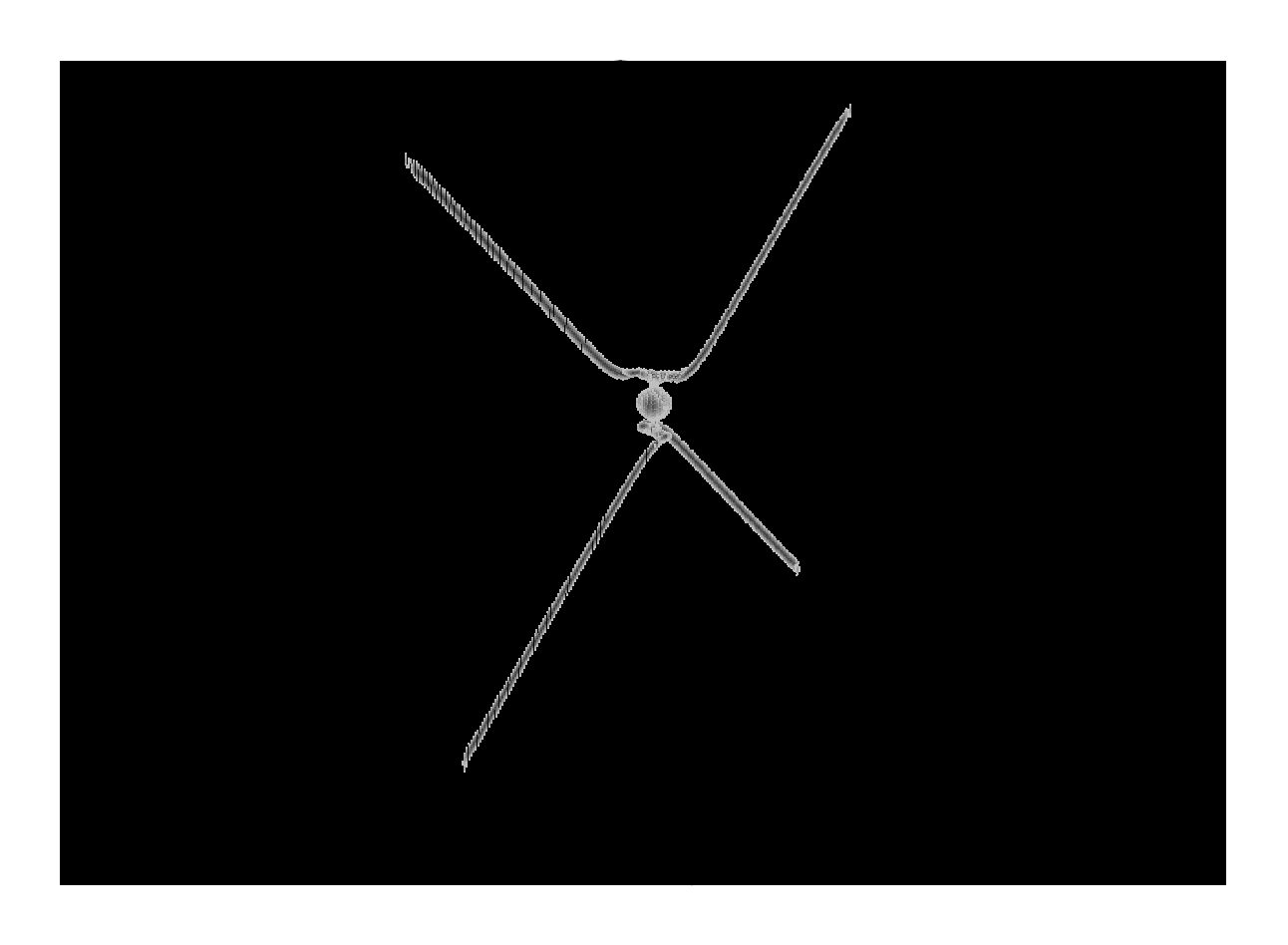} &
\includegraphics[height = 23mm,width=40mm]{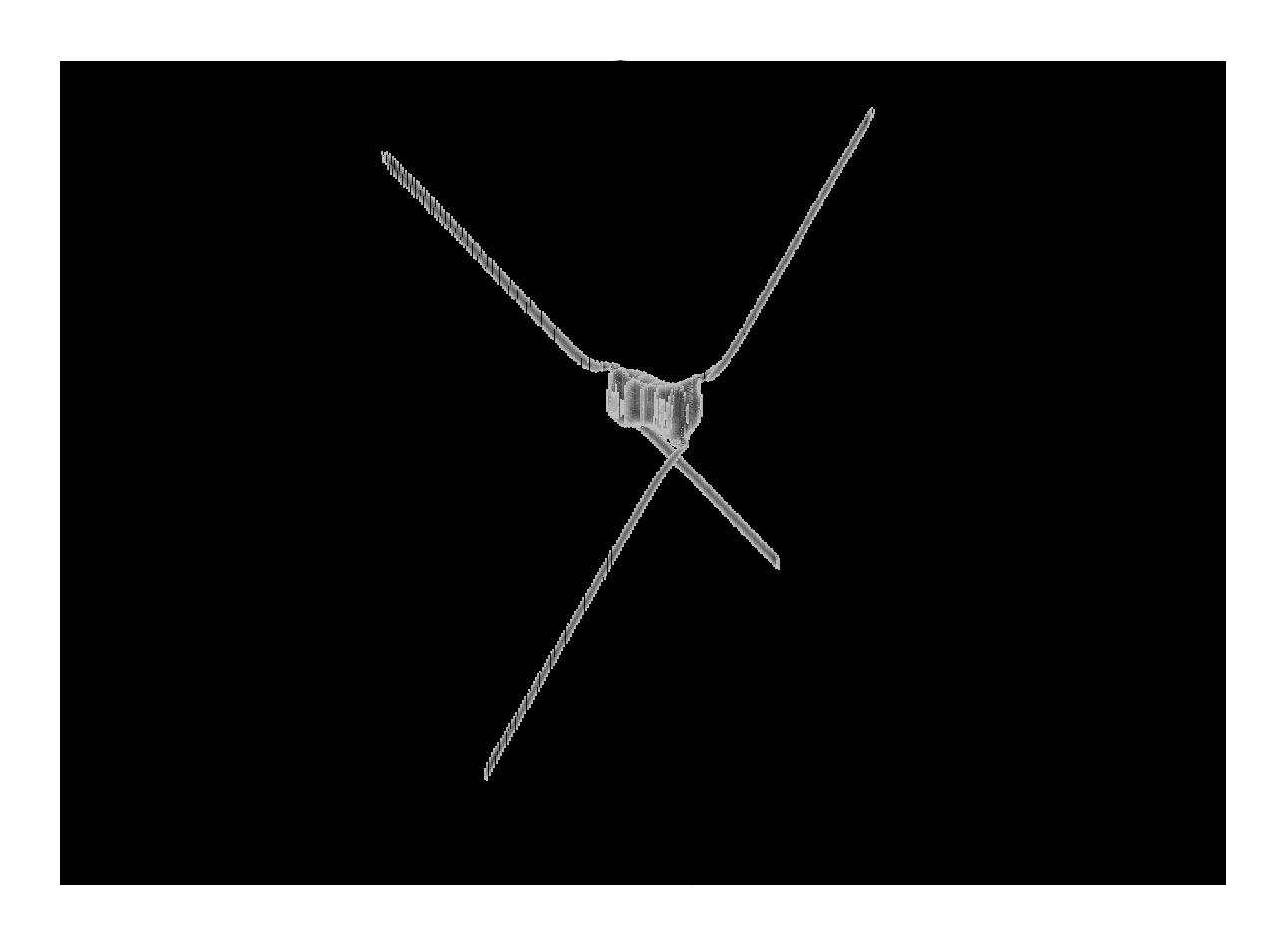} \\
\includegraphics[height = 23mm,width=40mm]{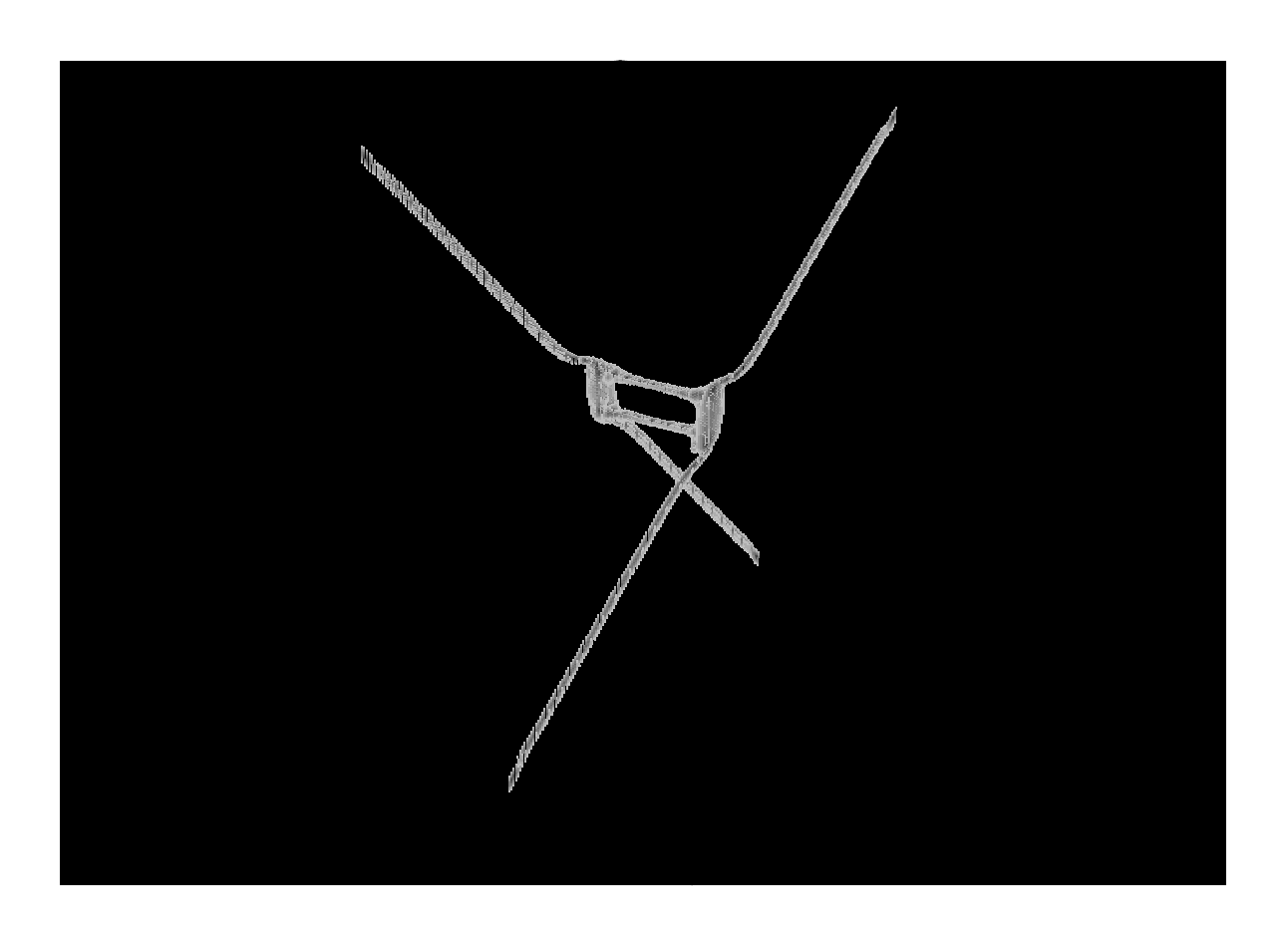} &
\includegraphics[height = 23mm,width=40mm]{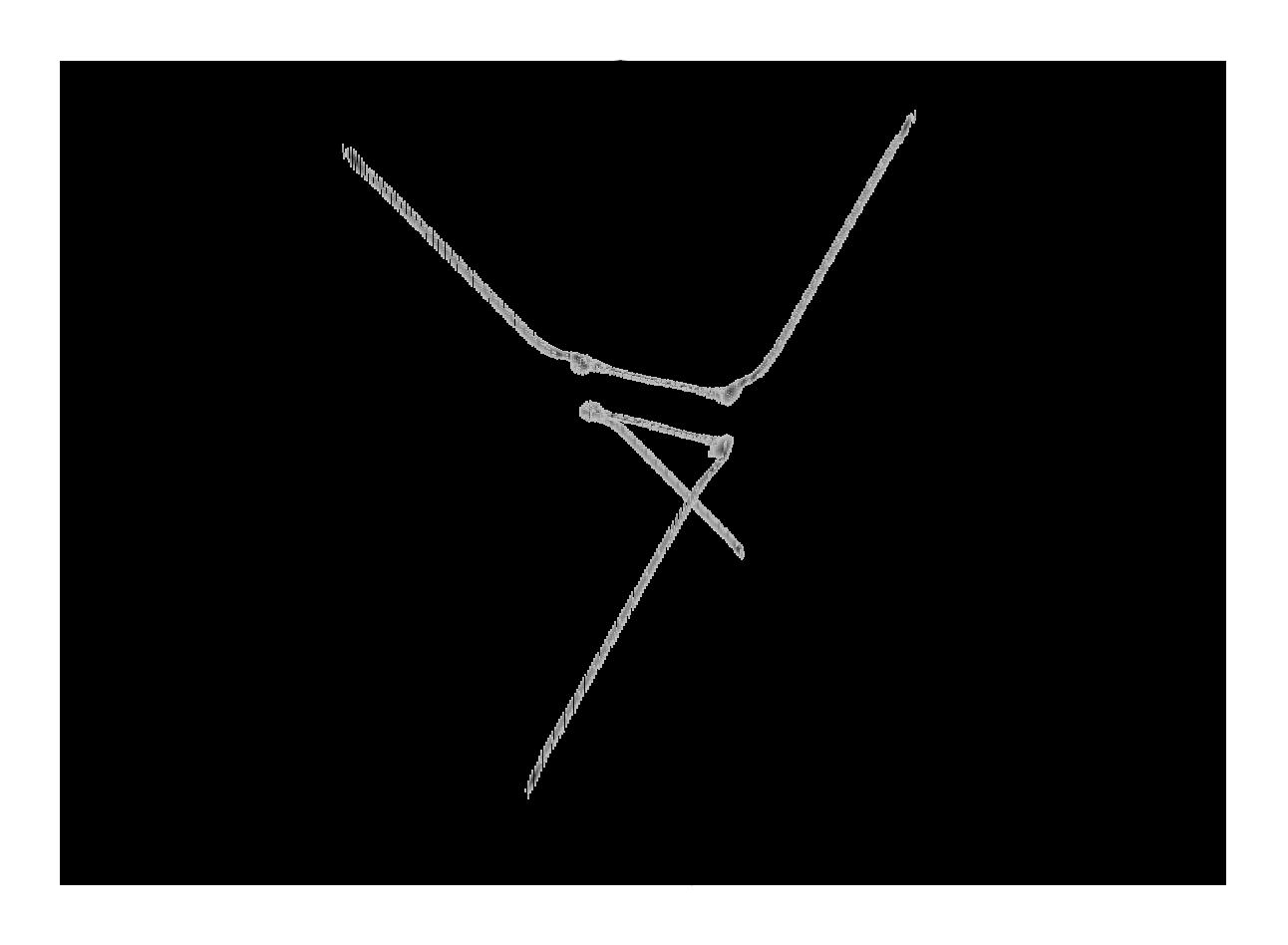} \\
\end{tabular}
\caption{After the first intercommutation a radiation blob
  emerges. The blob catches up with the bridge and is
  absorbed. However, before all radiation is absorbed, a loop seems to
  be formed. This ``loop'' is not topological, it breaks resulting in
  "Dracula's teeth". $(\beta = 4, \alpha = 122.7, v = 0.98)$
$t = 2, 4.5, 6.5, 8$.}
\label{blob_panels}
\end{minipage}
\end{figure}

$\bullet$ For $\beta \geq 16$, an expanding loop forms after a short
delay. If $v > v_c$, the loop catches up with the two original strings
and these reconnect again through the loop. This creates a highly
curved central region in each string, that move towards each other and
can sometimes mediate a third reconnection (see
fig. \ref{triple_panels}). At this point there are two almost
anti-aligned string segments and, if they are receding sufficiently
slowly, a fourth reconnection is possible (see
fig. \ref{quadruple_final}). This is the largest number of
intercommutations we have seen, and only for $\beta = 16$ (while triple
reconnections are generic).  On the other hand, if $v < v_c$ the
second reconnection does not take place because the string loop does
not catch up with the strings. In this case it will contract again and
eventually decay into radiation, sometimes after one or a few
oscillations.

$\bullet$ For $1< \beta \leq 8$ the energy isosurfaces look somewhat
similar, but they reveal a very different intermediate state. The
``loop'' in figure \ref{blob_panels} is just a blob of radiation with
no topological features: the (covariant) phase of the Higgs field
shows no net winding around the apparent vortex.
This is clearly visible in the third timestep, where the ``loop''
meets the string bridges, breaks and is absorbed -- a real vortex loop
would not be able to break. In this case the maximum number of
reconnections is two.

\section*{Discussion\label{Discussion}}


The interaction of the cores appears to play an important role in
interpreting the results.  In the type-II regime, parallel strings
repel. When strings approach each other, the interaction between the
magnetic cores produces a torque that tends to anti-align the string
segments, so the actual collision angle is higher than in the initial
configuration.  
However, due to this antialignment, the collision results in the
annihilation of a larger segment of string so in fact the strings
behave as if they that had collided with a {\it smaller} angle (which
is somewhat counterintuitive and is explained in \cite{us}). 
For collision angles approaching $180^o$ this leads to a different
prediction for the angular dependence of the critical velocity
(fig. \ref{vcrit_panels}.)

Our results appear to confirm the claim \cite{Putter} that the
critical speed goes down with increasing $\beta$, from $v_c \sim 0.96$
at $\beta = 1$ to $v_c \sim 0.86$ at $\beta = 64$, although this
reduction cannot go on indefinitely.  A crude (universal) lower bound
for $v_c$ follows from energy conservation: if the strings anti-align
locally and a portion $L$ of each string is annihilated, the maximum
energy available to the loop is $2L\gamma = 2 \pi R$, with $R$ the
maximum loop radius. If $R < L/2$ the second intercommutation cannot
take place, which happens for $v < \sqrt{1 - 4/\pi^2} \sim 0.77
$. These values of $v_c$ are to be contrasted with the average root
mean square velocity of the network, which has not yet been
investigated for type-II strings (known values range from $v_{rms} =
1/\sqrt 2 \sim 0.71$ for Nambu--Goto strings in flat space, to $~ 0.63
- 0.51$ in field theory simulations with $\beta = 1$ with cosmic
expansion \cite{MooreMartinsShellard,HindmarshStuckey}). So, double
intercommutations may be less rare than in type-I collisions, but they
are still rare events. However, once a kink train is formed its decay
time is comparable to that of a single kink, and because of its
microscopic size (it is only a few core widths in length) it is very
unlikely to be disrupted by intercommutation with another string
segment. We conclude that the small scale structure of strongly
type-II abelian Higgs string networks could be more clustered than the
predictions based on the Nambu-Goto approximation with $P=1$
\cite{KibbleCopeland,PolchinskiRocha,CopelandKibble_kinks}. The effect
on cosmological signatures is hard to predict, as multiple
intercommutations affect energy loss mechanisms in several, competing,
ways. First, if a string self-intersects, an even number of
intercommutations frustrates loop formation. Second, kinks tend to
suppress the formation of cusps. Third, kink trains would have a
stronger gravitational wave signal than single kinks and would give a
contribution to the stochastic background not considered
previously \cite{DamourVilenkin01,DamourVilenkin05,Siemens:2006vk,Olmez,Kawasaki}.
Finally, the contribution from reconnections is also enhanced in two
ways.  While we confirm that strongly type-II ANO strings always reconnect
($P=1$), an effective intercommutation probability $P \leq 1$ due to
multiple reconnections will still lead to denser networks and
therefore more reconnections. Most of these will be at low velocity,
where we expect strong anti-alignment, causing longer segments of
string to be annihilated into radiation than for $\beta \approx 1$
strings. So we would expect stronger and more frequent bursts of
radiation and cosmic rays than for other string types (lower $\beta$
and also superstrings) where reconnection bursts are always negligible
or subdominant \cite{JacksonSiemens}. Further work is needed to
understand these effects quantitatively, which affect the cosmological
bounds. If cosmic strings are ever observed, a crucial question will be
whether we can infer their microscopic structure from the
observations, in particular whether they are {\it just} topological
defects or the first evidence of superstring theory in the sky. The
results presented here suggest that, even in the relatively simple
case of ANO strings, 
the small scale structure may still hold surprises.

\begin{acknowledgments}
  Work supported by the Netherlands Organization for Scientific Research
  (NWO) under the VICI programme.
\end{acknowledgments}

\bibliography{bib}

\end{document}